\documentclass[twocolumn,amsmath,amssymb,superscriptaddress,notitlepage]{revtex4-1}
\usepackage{graphics}
\usepackage{amsmath,bm}
\usepackage{dcolumn}
\usepackage{natbib}
\usepackage{epstopdf}
\usepackage{float}
\usepackage{graphicx}
\usepackage{epsfig}
\usepackage[pdfstartview=FitH]{hyperref}
\usepackage{color}
\usepackage{appendix}
\usepackage{ulem}

\newcommand{\rr}{{\bf r}}

\begin{document}
\title{Three-component fractional quantum Hall effect in topological flat bands}
\author{Tian-Sheng Zeng}
\affiliation{Department of Physics, College of Physical Science and Technology, Xiamen University, Xiamen 361005, China}
\date{\today}
\begin{abstract}
We study the many-body ground states of three-component quantum particles in two prototypical topological lattice models under strong intercomponent and intracomponent repulsions. At band filling $\nu=3/4$ for hardcore bosons, we demonstrate the emergence of three-component fractional quantum Hall (FQH) effect characterized by the $\mathbf{K}$ matrix, through exact diagonalization study of four-fold quasidegenerate ground states with a robust spectrum gap and the combined density-matrix renormalization group calculation of fractional drag charge pumping. Further we formulate the topological characterization of FQH states of three-component Bose-Fermi mixtures at various fillings by the $\mathbf{K}$ matrix. At last we discuss the possible generalization of our approach to identify non-Abelian three-component spin-singlet FQH states.
\end{abstract}
\maketitle

\section{Introduction}

Since Halperin's classic proposal of two-component quantum Hall effect~\cite{Halperin1983}, the occurence of multicomponent fractional quantum Hall (FQH) effects effects gradually attracted attention when an internal degree of freedom
(i.e., spin and layer) was included~\cite{Wen1992,Chakraborty1987,Yoshioka1988,Yoshioka1989,He1991,He1993,Seidel2008}. Subsequently, after the experimental discovery of a rich class of approximately SU(4) symmetric FQH effects in graphene sheets with two-fold spin and two-fold valley degrees of freedom~\cite{Bolotin2009,Dean2011}, many theoretical investigations proposed a generalization of Halperin's wave functions to four-component cases, such as SU(4) composite-fermion construction~\cite{Toke2007,Balram2015} and SU(4) generalization of Halperin's trial wavefunction~\cite{Goerbig2007,Gail2008}. Nevertheless multicomponent quantum Hall systems offer us a new scope for seeking topological phases that have no analog in one-component systems.~\cite{Wen2017}. Indeed, several exotic interlayer-correlated FQH effects have been experimentally observed in two separated parallel graphene layers~\cite{Liu2019,Li2019}. Among these multicomponent FQH systems, different intercomponent and intracomponent correlations play the vital role. However most of the prior active studies focus on exploring even-component cases including two-component bosonic quantum Hall effect like Halperin $(221)$ FQH effect~\cite{Grass2012,Furukawa2012,Wu2013,Grass2014}, bosonic integer quantum Hall effect~\cite{Senthil2013,Furukawa2013,Regnault2013,Wu2013,Grass2014} and even non-Abelian spin-singlet quantum Hall effects (dubbed as NASS states)~\cite{Ardonne1999,Reijnders2002,Barkeshli2010}. To date, very little knowledge about three-component FQH effects are acquired, because clear-cut examples are relatively rare~\cite{Sodemann2014,Wu2015}.

In a parallel route, the exciting arise of topological Chern bands (such as realization of Haldane Chern insulator~\cite{Zhao2024} and observation of fractional quantum anomalous Hall effect with an internal spin-valley locking degree of freedom in moir\'e superlattices~\cite{Park2023,Kang2024}) in analogy to two-dimensional Landau levels endows us a prime candidate for studying multicomponent quantum Hall effects. In particular, it is numerically suggested that FQH effects at specific band fillings $\nu=1/(C+1)$ (for spinless hardcore bosons) and $\nu=1/(2C+1)$ (for spinless fermions) in topological flat bands with high Chern numbers $C>1$, are described as color-entangled lattice versions of $C$-component FQH effects where $C$ serves as an internal degree of freedom~\cite{LBFL2012,Wang2012r,Yang2012,Sterdyniak2013,YLWu2013,YLWu2014,YHWu2015,Behrmann2016,Andrews2018,Dong2023}. However, a closely related drawback is that it is hard to directly discern mutual particle correlations contributed by each component in these lattice models~\cite{Wang2023}. Alternatively, we may consider multicomponent quantum particles loaded in topological flat bands with Chern number $C=1$ as the substitute. The merit of these models is that now we can distinguish intercomponent and intracomponent correlations in a designed many-body Hamiltonian. Extensive numerical calculations have demonstrated FQH effects of multicomponent quantum particles in topological flat bands with unit Chern number~\cite{Zeng2017,Zeng2018,Zeng2019,Zeng2020,Zeng2022a,Zeng2022,Zeng2023}, which are characterized by the $\mathbf{K}$-matrix within the framework of the Chern-Simons gauge-field theory~\cite{Wen1992a,Wen1992b,Blok1990a,Blok1990b,Blok1991}. Moreover, the $\mathbf{K}$-matrix also solely determines the modular matrix which encodes the statistics of anyonic excitations for multicomponent FQH states~\cite{Hu2023}. Interestingly, recent studies also suggest that two-component Bose-Fermi mixtures in topological bands can host exotic chiral topological phases~\cite{Wu2019,Zeng2021}. A spectacular question is whether new kinds of fascinating FQH effects exist in three-component systems or not.

Here, we aim at searching for different possible exotic three-component FQH effects at various filling factors in topological lattice models with the lowest band carrying unit Chern number $C=1$, and then further identify their topological properties. This paper is organized as follows. In Sec.~\ref{model}, we introduce the microscopic interacting Hamiltonian of three-component quantum particles loaded on two prototypical topological lattice models, and give a description of our numerical methods. In Sec.~\ref{Halperin}, we discuss the FQH states of three-component bosons, and present numerical results of their topological characterization based on the $\mathbf{K}$ matrix by exact diagonalization and density-matrix renormalization group calculations. In Sec.~\ref{Mixture}, we discuss the identification of more exotic FQH effects of three-component Bose-Fermi mixtures based on the $\mathbf{K}$ matrix; following that, we present our numerical results for a mixture of two-component bosons and one-component fermions in Sec.~\ref{b2f1}, and a mixture of one-component bosons and two-component fermions in Sec.~\ref{b1f2}. In Sec.~\ref{nonabelian}, we discuss the non-Abelian spin-singlet FQH effect. Finally, we present a summary and discussion in Sec.~\ref{summary}, and discuss the relationship between the charge transfer and the quasiparticle charge.

\section{Model and Method}\label{model}

In this section, we begin with the description of three-component quantum particles (either bosons or fermions) with strong intercomponent and intracomponent repulsions in topological flat bands, and introduce our density-matrix renormalization group (DMRG) and exact diagonalization (ED) simulations. We study two different prototypical topological lattice models, i.e. $\pi$-flux checkerboard (CB) lattice~\cite{Sun2011} and Haldane-honeycomb lattice~\cite{Wang2011}. The model Hamiltonian is formulated as
\begin{align}
  H_{CB}=&\!\sum_{\sigma}\!\Big[-t\!\!\sum_{\langle\rr,\rr'\rangle}\!e^{i\phi_{\rr'\rr}}c_{\rr',\sigma}^{\dag}c_{\rr,\sigma}
  -\!\!\!\!\sum_{\langle\langle\rr,\rr'\rangle\rangle}\!\!\!t_{\rr,\rr'}'c_{\rr',\sigma}^{\dag}c_{\rr,\sigma}\nonumber\\
  &-t''\!\!\!\sum_{\langle\langle\langle\rr,\rr'\rangle\rangle\rangle}\!\!\!\! c_{\rr',\sigma}^{\dag}c_{\rr,\sigma}+H.c.\Big]+V_{int},\label{cbl}\\
  H_{HC}=&\!\sum_{\sigma}\!\Big[-t\!\!\sum_{\langle\rr,\rr'\rangle}\!\! c_{\rr',\sigma}^{\dag}c_{\rr,\sigma}-t'\!\!\sum_{\langle\langle\rr,\rr'\rangle\rangle}\!\!e^{i\phi_{\rr'\rr}}c_{\rr',\sigma}^{\dag}c_{\rr,\sigma}\nonumber\\
  &-t''\!\!\sum_{\langle\langle\langle\rr,\rr'\rangle\rangle\rangle}\!\!\!\! c_{\rr',\sigma}^{\dag}c_{\rr,\sigma}+H.c.\Big]+V_{int},\label{hcl}
\end{align}
with the many-body interaction given by
\begin{align}
  V_{int}=U\sum_{\sigma\neq\sigma'}\sum_{\rr}n_{\rr,\sigma}n_{\rr,\sigma'}+\sum_{\sigma}V_{\sigma}\sum_{\langle\rr,\rr'\rangle}n_{\rr',\sigma}n_{\rr,\sigma}.\label{interact}
\end{align}
Here $\langle\ldots\rangle$,$\langle\langle\ldots\rangle\rangle$ and $\langle\langle\langle\ldots\rangle\rangle\rangle$ denote the nearest-neighbor, next-nearest-neighbor, and next-next-nearest-neighbor pairs of sites, and we choose the corresponding tunnel couplings $t'=0.3t,t''=-0.2t,\phi=\pi/4$ for checkerboard lattice and $t'=0.6t,t''=-0.58t,\phi=2\pi/5$ for honeycomb lattice respectively~\cite{Wang2011}, whose lowest band is rather flat and host a Chern number $C=1$ (See the Appendix~\ref{band} for lattice geometry). The subscript labels $\sigma=0,\pm1$ are used as the pseudospin indices to distinguish different components, $c_{\rr,\sigma}^{\dag}$ is the particle creation operator of the spin-$\sigma$ component at site $\rr$, $n_{\rr,\sigma}=c_{\rr,\sigma}^{\dag}c_{\rr,\sigma}$ is the particle number operator of the spin-$\sigma$ component at site $\rr$. $U$ is the strength of the onsite intercomponent repulsion, and $V_{\sigma}$ is the intracomponent repulsion strength between the nearest-neighbor pairs of the spin-$\sigma$ component.

In our numerical calculations, we fix the global particle number $N_{\sigma}$ for each component (that means our models are constrained by $U(1)\times U(1)\times U(1)$ symmetry), and take the finite lattice systems with $N_x\times N_y$ unit cells (then $N_s=2\times N_x\times N_y$ is the total lattice sites). For small system sizes with periodic torus geometries, we utilize the ED study with the energy states labeled by the total momentum $K=(K_x,K_y)$ in units of $(2\pi/N_x,2\pi/N_y)$ in the Brillouin zone. For large system sizes, we exploit finite DMRG on the cylindrical ladder geometry with finite width $N_y$ and finite length $N_x$. We limit the largest cylinder width $N_y=4$ and length $N_x=42$ and keep the maximal bond dimension up to $M=4000$, and the maximal discarded truncation error is less than $3\times10^{-5}$. The boundary condition of cylinder ladder is open in the $x$ direction and periodic in the $y$ direction. To avoid the local minimum state, we choose different random initial states with the sweep number more than 30 to get the most convergent ground state.

\section{Three-component Bosonic Fractional Quantum Hall Effect}\label{Halperin}

\begin{figure}[b]
  \includegraphics[height=1.75in,width=3.35in]{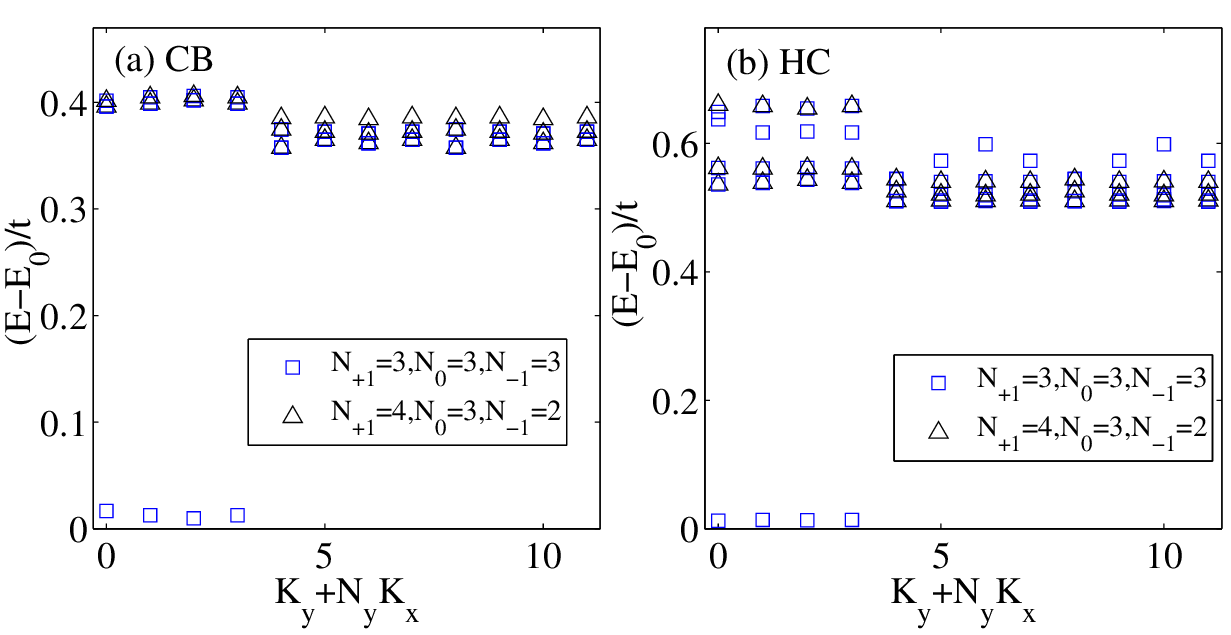}
  \caption{\label{energy}(Color online) Numerical ED results for the low energy spectrum of three-component hardcore bosons at filling $\nu=3/4$ with $U=\infty,V_{\sigma}=0$. The system size $N_s=2\times3\times4$ is used for (a) topological checkerboard lattice and (b) topological honeycomb lattice. The lowest four energy levels in each momentum sector are shown.
  }
\end{figure}

\begin{table*}[t]
\setlength{\tabcolsep}{0.3cm}
\caption{\label{counting} The total momentum counting in one-dimensional lattice of orbitals of momentum $k_y+N_yk_x$ for three-component hardcore bosons at filling $\nu=3/4$ with $N_x=3,N_y=4$.}
\begin{tabular}{|c|c|c|c|c|c|c|c|c|c|c|c|c|c|c|}
\hline
\hline
$k_x$ & \multicolumn{4}{|c|}{0} & \multicolumn{4}{|c|}{1} & \multicolumn{4}{|c|}{2}  & \multicolumn{2}{|c|}{Total Momentum $(K_x,K_y)$} \\
\hline
$k_y$ & 0 & 1 & 2 & 3 & 0 & 1 & 2 & 3 & 0 & 1 & 2 & 3 & $K_x$ & $K_y$ \\
\hline
State 1 & $X$ & $X$ & $X$ & 0 & $X$ & $X$ & $X$ & 0 & $X$ & $X$ & $X$ & 0 & 9(mod 3)=0 & 9(mod 4)=1 \\
\hline
State 2 & $X$ & $X$ & 0 & $X$ & $X$ & $X$ & 0 & $X$ & $X$ & $X$ & 0 & $X$ & 9(mod 3)=0 & 12(mod 4)=0 \\
\hline
State 3 & $X$ & 0 & $X$ & $X$ & $X$ & 0 & $X$ & $X$ & $X$ & 0 & $X$ & $X$ & 9(mod 3)=0 & 15(mod 4)=3 \\
\hline
State 4 & 0 & $X$ & $X$ & $X$ & 0 & $X$ & $X$ & $X$ & 0 & $X$ & $X$ & $X$ & 9(mod 3)=0 & 18(mod 4)=2 \\
\hline
\end{tabular}
\end{table*}

In two-dimensional quantum Hall systems, an ingenious generalization of two-component Halperin's variational wavefunction to higher three-component cases, can be written as (apart from a product factor):
\begin{align}
 \Psi\propto \prod_{i<j,\sigma}(z_i^{\sigma}-z_j^{\sigma})^{m}
 \prod_{i,j,\sigma\neq\sigma'}(z_i^{\sigma}-z_j^{\sigma'})^{n},\nonumber
\end{align}
whose topological order could be classified by the $\mathbf{K}$-matrix
\begin{align}
  \mathbf{K}=\begin{pmatrix}
m & n & n\\
n & m & n\\
n & n & m\\
\end{pmatrix}.\label{kmatrix}
\end{align}
For simplicity we take complex coordinates $z_j^{\sigma}=x_j^{\sigma}+iy_j^{\sigma}$ of the $j$-th particles with spin-$\sigma$ ($j=1,2,\cdots,N_{\sigma}$), and suppose the positive values $m,n>0$. From the above equation, we can derive that the first spin-$\sigma$ particle $z_1^{\sigma}$ hosts the highest orbital angular momentum (also the number of orbital states) $N_{\phi}=m(N_{\sigma}-1)+n\sum_{\sigma'\neq\sigma}N_{\sigma'}$ (in units of $\hbar$) in the lowest Landau level under symmetric gauge. Due to cyclic permutation symmetry of the subscript labels, $N_{\sigma}=N_{\sigma'}$. For large particle numbers $N_{\sigma}\gg1$, we obtain the spin-$\sigma$ filling $\nu_{\sigma}=N_{\sigma}/N_{\phi}=1/(m+2n)$ and the total particle filling $\nu=\sum_{\sigma}\nu_{\sigma}=3/(m+2n)$. In connection to our lattice models, we fix the particle fillings of the lowest Chern band at $\nu=\sum_{\sigma}N_{\sigma}/(N_xN_y)$ to simulate the FQH states. Here we consider the FQH states of three-component hardcore bosons at filling $\nu=3/4$ under strong onsite repulsions $U/t\gg1$ without nearest-neighbor interaction $V_{\sigma}=0$. In the following part we shall characterize the $\mathbf{K}$ matrix from topological degeneracy, topologically invariant Chern number, and fractional charge pumping.

We take the commensurate finite periodic lattice $N_s=2\times3\times4$ in the ED study with the total particle number $N=N_{+1}+N_0+N_{-1}=9$. Here we set $U/t=\infty$ (no more than one particle per site), and the largest reduced Hilbert space is still up to $1.8\times10^{8}$ in each momentum sector, which would cost around 1Tb of the computational memory. As shown in Fig.~\ref{energy}, we plot the low energy spectrum in typical charge sectors $N_{+1},N_0,N_{-1}$ for different lattice models, and find that the many-body ground states are four-fold degenerate, with a large robust gap separated from the higher excited levels. Moreover, they fall into the charge sector $N_{+1}=N_0=N_{-1}$, which are of SU(3) spin-singlet nature.

\begin{figure}[b]
  \includegraphics[height=1.75in,width=3.35in]{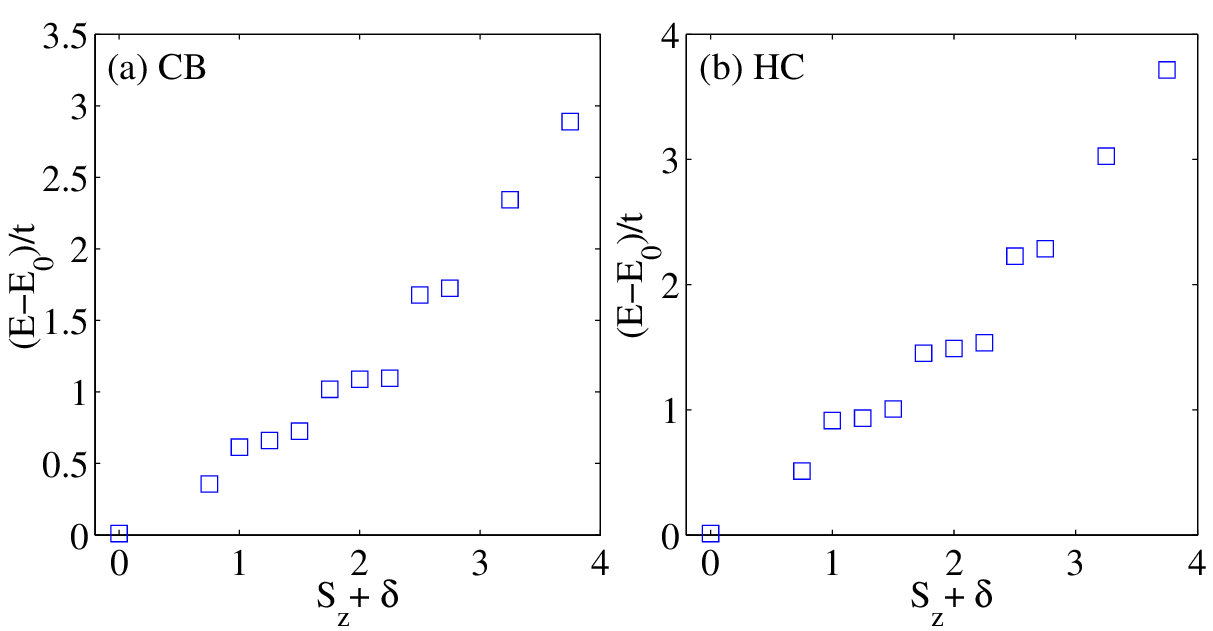}
  \caption{\label{spin} (Color online) Numerical ED results for the lowest energy level of three-component hardcore bosons at $\nu=3/4$ with $U=\infty,V_{\sigma}=0$ as a function of spin polarization $S_z+\delta$ for (a) topological checkerboard lattice and (b) topological honeycomb lattice. The system size $N_s=2\times3\times4$.}
\end{figure}

\begin{figure}[b]
  \includegraphics[height=2.45in,width=3.3in]{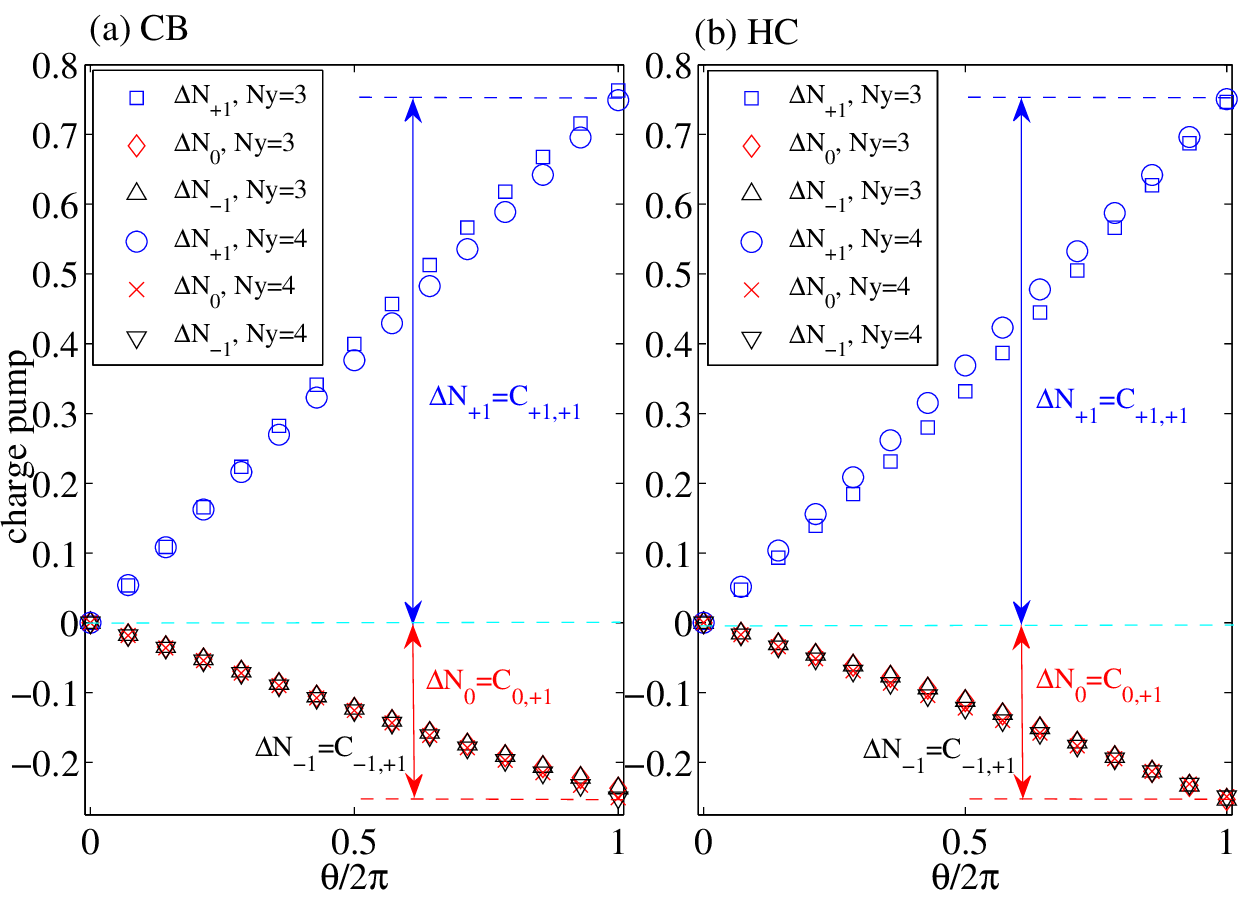}
  \caption{\label{pump} (Color online) Fractional charge transfers for three-component hardcore bosons at $\nu=3/4$ with $U=\infty,V_{\sigma}/t=0,$ on the cylinder lattice under the insertion of flux quantum $\theta_{+1}^{y}=\theta,\theta_{0}^{y}=\theta_{-1}^{y}=0$ for two different topological models: (a) checkerboard model and (b) Haldane-honeycomb model. Here finite DMRG is used with different cylinder widths $N_y=3,4$, keeping the cylinder length up to 40.}
\end{figure}

To further unravel the spin-singlet nature, we focus on the charge sectors $N_{+1}\geq N_0\geq N_{-1}$, and define the spin polarization of three-component quantum particles as $S_z+\delta$ with magnetization $S_z=\sum_{\sigma}\sigma N_{\sigma}/3$ and relative magnetization $\delta=(N_{+1}-N_0)/(N_xN_y)$. As illustrated in Figs.~\ref{spin}(a) and~\ref{spin}(b), we plot the ground state energy of different spin polarizations, and find that the ground state energy satisfies two characteristics: (i) $E_0(S_z,\delta)<E_0(S_z',\delta)$ for $S_z<S_z'$ and (ii) $E_0(S_z,\delta)<E_0(S_z,\delta')$ for $\delta<\delta'$. Therefore we establish the fact that the lowest energy level always falls into the spin-singlet sector $S_z=0,\delta=0$ (namely $N_{+1}=N_0=N_{-1}$). Also we note that according to the squeezing rule~\cite{Ardonne2011}, the four root configurations of SU(3) spin-singlet FQH states at $\nu=3/4$ are $(XXX0)$ and the other translational invariant partners $(XX0X),(X0XX),(0XXX)$, where $XXX$ stands for a superposition of $|+1\rangle|0\rangle|-1\rangle$ and its five permutation partners $\sigma\leftrightarrow\sigma'$. And the total momentum counting by the generalized Pauli principle for the four-fold degenerate ground states for $N=9$ particles in a $N_s=2\times3\times4$ lattice is indicated in Table~\ref{counting}, and their total momenta are located at $(K_x,K_y)=(0,i)$ ($i=0,1,2,3$), consistent with our numerical analysis.

Due to the huge computational memory cost, to overcome the computational difficulty of ED study in calculating topological Chern number, instead we use finite DMRG, and calculate the fractional charge pump to underscore the topological nature, where quantized charge pumping via an adiabatically periodic perturbation is a hallmark of the two-dimensional quantum Hall effect~\cite{Laughlin1981}, which is connected to the Hall conductance (Chern number). Remarkably, for the intercomponent Chern number $C_{\sigma\sigma'}$, it is shown that a quantized drag Hall conductance can be obtained by threading one flux quantum (namely, the twisted angle $\theta_{\sigma}^{\alpha}$ changes from zero to $2\pi$~\cite{Gong2014}).

\begin{figure}[t]
  \includegraphics[height=1.8in,width=3.2in]{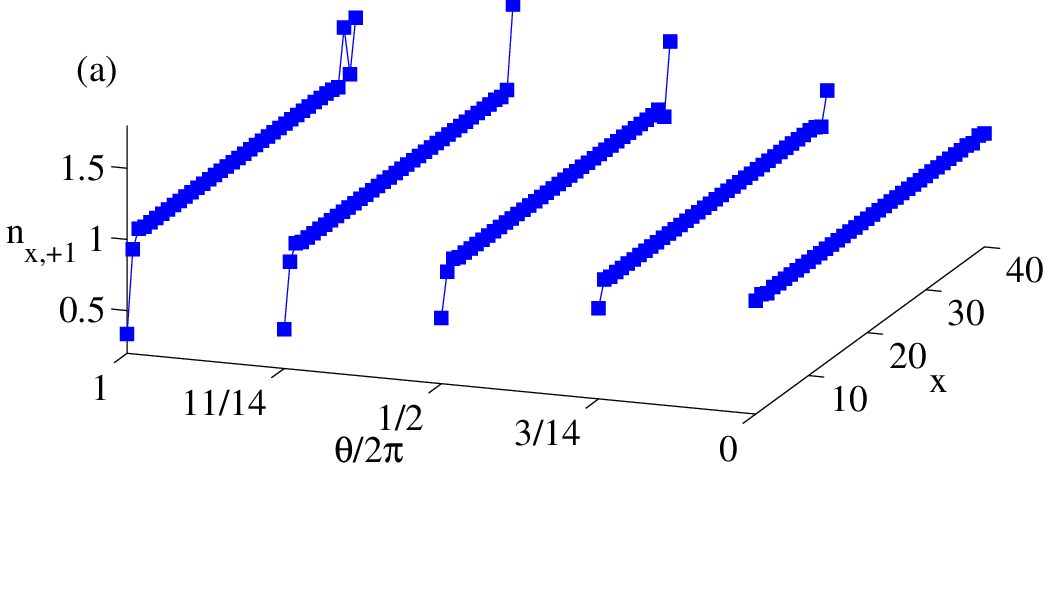}
  \includegraphics[height=1.75in,width=3.2in]{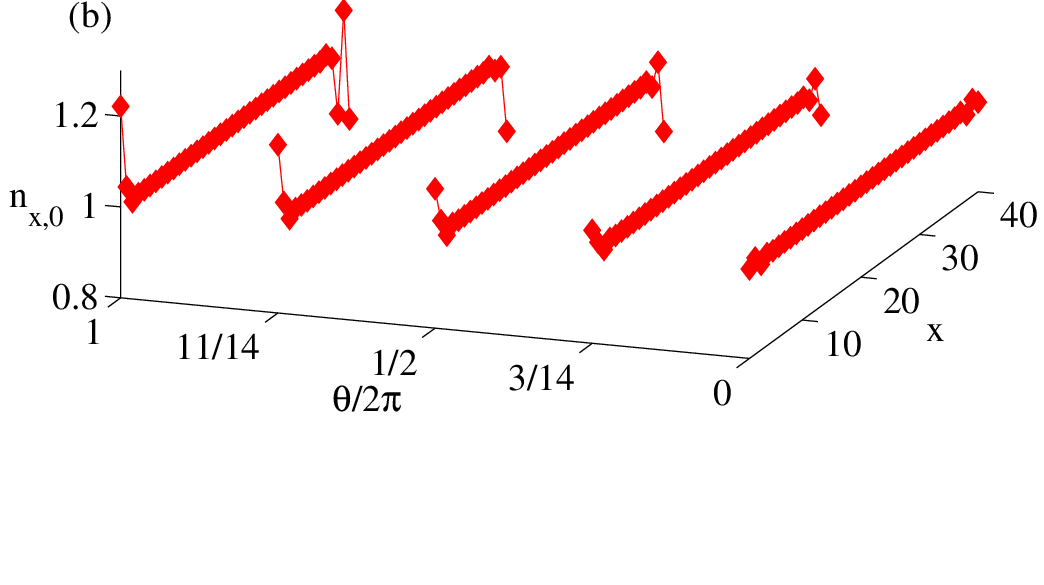}
  \includegraphics[height=1.4in,width=3.2in]{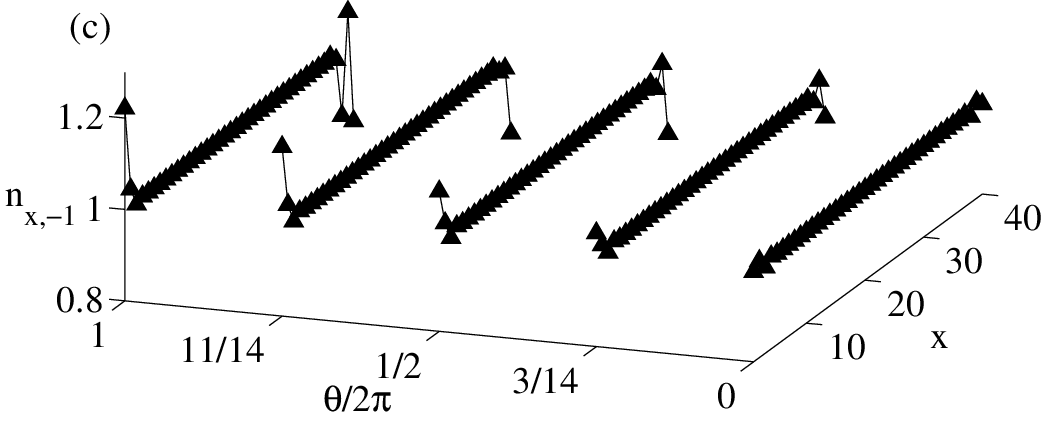}
  \caption{\label{edge} (Color online) Real-space configurations of particle accumulations $n_{x,\sigma}=\sum_{y}n_{\rr,\sigma}$ in each column for three-component hardcore bosons at $\nu=3/4$ with $U=\infty,V_{\sigma}/t=0,$ on the cylinder lattice under the insertion of flux quantum $\theta_{+1}^{y}=\theta,\theta_{0}^{y}=\theta_{-1}^{y}=0$ for topological Haldane-honeycomb model. Here finite DMRG is used with cylinder width $N_y=4$ and length $N_x=40$.}
\end{figure}

To simulate this physical effect, we utilize finite DMRG to calculate the charge pumping on finite cylinder ladders under the adiabatic insertion of one flux quantum. Numerically we focus on the charge sector $N_{+1}=N_0=N_{-1}$, and cut the finite cylinder into equal left-half and right-half parts along the $x$ direction. As the twisted angle $\theta_{\sigma'}^{y}$ is changed, we define the total particle number of the spin-$\sigma$ component in the right-half part as $N_{\sigma}^{R}(\theta_{\sigma'}^{y})=\sum_{\rr}n_{\rr,\sigma}^R$ where $n_{\rr,\sigma}^R$ is the occupation number on lattice sites of the right-half part. By tracing the evolution of $N_{\sigma}^{R}(\theta_{\sigma'}^{y})$ as a function of $\theta_{\sigma'}^{y}$, we obtain the net charge transfers for the spin-$\sigma$ particle from the left-half to the right-half of the cylinder. As illustrated in Figs.~\ref{pump}(a) and~\ref{pump}(b), we plot the evolution of the charge transfer under the insertion of one flux quantum $\theta_{+1}^{y}=\theta,\theta_{0}^{y}=\theta_{-1}^{y}=0$ for two different types of topological lattice models, and find the fractionally quantized charge pumpings
\begin{align}
  &\Delta N_{+1}=N_{+1}^R(2\pi)-N_{+1}^R(0)\simeq C_{+1,+1}=\frac{3}{4}, \nonumber\\
  &\Delta N_{0}=N_{0}^R(2\pi)-N_{0}^R(0)\simeq C_{0,+1}=-\frac{1}{4}, \nonumber\\
  &\Delta N_{-1}=N_{-1}^R(2\pi)-N_{-1}^R(0)\simeq C_{-1,+1}=-\frac{1}{4}. \nonumber
\end{align}
We also measure the corresponding real-space configuration of particle accumulation along the cylinder length, defined as $n_{x,\sigma}=\sum_{y}n_{\rr,\sigma}$ (the summation is done over all the $2N_y$ sites in each column $x$). As shown in Figs.~\ref{edge}(a-c), we find nontrivial edge charge polarization with the increase of $\theta_{\sigma'}^y$: the particles of spin-$\sigma$ accumulate at one edge of the cylinder while they diminish at the other edge of the cylinder, which is equivalent to the charge pumping $C_{\sigma,\sigma'}$ from one edge to the other edge in the above discussion. Because of the permutation symmetry of pseudospin $c_{\rr,\sigma}\rightleftharpoons c_{\rr,\sigma'}$ in the model Hamiltonian, we obtain the other Chern numbers $C_{0,0}=C_{-1,-1}=C_{+1,+1}$ and $C_{\sigma,\sigma'}=C_{\sigma',\sigma}$. Then the topological $\mathbf{K}$-matrix classification is given by the inverse of the Chern number matrix
\begin{align}
  \mathbf{K}=\begin{pmatrix}
C_{+1,+1} & C_{+1,0} & C_{+1,-1}\\
C_{0,+1} & C_{0,0} & C_{0,-1}\\
C_{-1,+1} & C_{-1,0} & C_{-1,-1}\\
\end{pmatrix}^{-1}
=\begin{pmatrix}
2 & 1 & 1\\
1 & 2 & 1\\
1 & 1 & 2\\
\end{pmatrix}.
\end{align}
In turn, the determinant of this $\mathbf{K}$-matrix $\det|\mathbf{K}|=4$ matches with the topological degeneracy of the ground state manifold in our ED study, as required.

\section{Fractional Quantum Hall Effect of three-component Bose-Fermi Mixtures}\label{Mixture}

Following the last section, we turn to analyze the emergence of multicomponent FQH effect of three-component Bose-Fermi mixtures, whose topological order can be classified by the symmetric integer-valued $\mathbf{K}$-matrix
\begin{align}
  \mathbf{K}=\begin{pmatrix}
m_{+1} & n_1 & n_2\\
n_1 & m_0 & n_3\\
n_2 & n_3 & m_{-1}\\
\end{pmatrix}.\label{bfmatrix}
\end{align}
Due to individual quantum statistics of bosons and fermions, the diagonal elements of the $\mathbf{K}$-matrix are unequal $m_{\sigma}\neq m_{\sigma'}$. Similar to what we did in Sec.~\ref{Halperin}, we continue to apply finite DMRG to calculate the charge pumping under the adiabatic insertion of one flux quantum and construct the Chern number matrix.

\subsection{Mixture of two-component bosons and one-component fermions}\label{b2f1}

\begin{figure}[t]
  \includegraphics[height=2.5in,width=3.35in]{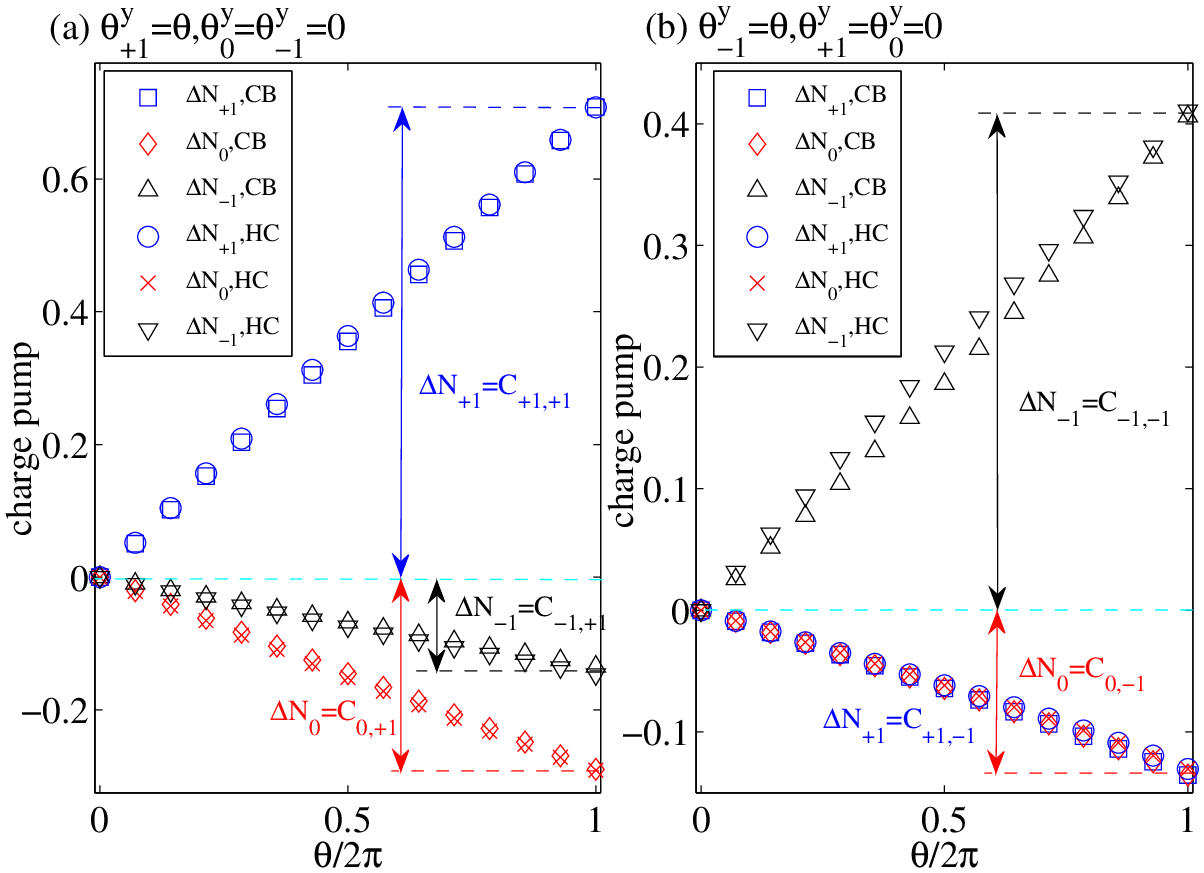}
  \caption{\label{pump223} (Color online) Fractional charge transfers for two-component hardcore bosons at $\nu_{+1}=\nu_{0}=2/7$ and one-component fermions at $\nu_{-1}=1/7$ with $U=\infty,V_{-1}/t=100,V_{+1}/t=V_{0}/t=0,$ on the cylinder lattice for two different topological models under the insertion of two types of flux quantum: (a) $\theta_{+1}^{y}=\theta,\theta_{0}^{y}=\theta_{-1}^{y}=0$ and (b) $\theta_{-1}^{y}=\theta,\theta_{+1}^{y}=\theta_{0}^{y}=0$. Here finite DMRG is used with cylinder width $N_y=4$ and length $N_x=42$.}
\end{figure}

We first investigate a mixture of two-component bosons and one-component fermions in the model Hamiltonian, where the particles of spin-$\sigma=+1,0$ component are taken as hardcore bosons while the particles of spin-$\sigma=-1$ component are treated as spinless fermions. We consider the FQH states of two-component bosons at fillings $\nu_{+1}=\nu_0=2/7$ and one-component fermions at filling $\nu_{-1}=1/7$ with strong onsite repulsions $U/t\gg1$ among different components and only intracomponent nearest-neighboring repulsion between fermions (namely $V_{-1}/t\gg1$ and $V_{+1}=V_0=0$).

As illustrated in Figs.~\ref{pump223}(a) and~\ref{pump223}(b) for different types of flux insertion in different topological lattice models, we calculate the charge pumpings $\Delta N_{\sigma}=N_{\sigma}^{R}(\theta_{\sigma'}^{y}=2\pi)-N_{\sigma}^{R}(\theta_{\sigma'}^{y}=0)$ and find that (i) for the flux insertion of spin-$\sigma=+1$ bosons $\theta_{+1}^y=\theta,\theta_0^y=\theta_{-1}^y=0,\theta\subseteq[0,2\pi]$,
\begin{align}
  &\Delta N_{+1}=N_{+1}^R(2\pi)-N_{+1}^R(0)\simeq C_{+1,+1}=\frac{5}{7}, \nonumber\\
  &\Delta N_{0}=N_{0}^R(2\pi)-N_{0}^R(0)\simeq C_{0,+1}=-\frac{2}{7}, \nonumber\\
  &\Delta N_{-1}=N_{-1}^R(2\pi)-N_{-1}^R(0)\simeq C_{-1,+1}=-\frac{1}{7}. \nonumber
\end{align}
and (ii) for the flux insertion of spin-$\sigma=-1$ fermions $\theta_{+1}^y=\theta_0^y=0,\theta_{-1}^y=\theta,\theta\subseteq[0,2\pi]$,
\begin{align}
  &\Delta N_{+1}=N_{+1}^R(2\pi)-N_{+1}^R(0)\simeq C_{+1,-1}=-\frac{1}{7}, \nonumber\\
  &\Delta N_{0}=N_{0}^R(2\pi)-N_{0}^R(0)\simeq C_{0,-1}=-\frac{1}{7}, \nonumber\\
  &\Delta N_{-1}=N_{-1}^R(2\pi)-N_{-1}^R(0)\simeq C_{-1,-1}=\frac{3}{7}. \nonumber
\end{align}
From the permutation symmetry $c_{\rr,+1}\leftrightarrow c_{\rr,0}$, the diagonal and off-diagonal elements $C_{0,0}=C_{+1,+1},C_{0,-1}=C_{+1,-1}$. Now we can obtain the $\mathbf{K}$-matrix from the inverse of the Chern number matrix, namely
\begin{align}
  \mathbf{K}=\mathbf{C}^{-1}=\begin{pmatrix}
2 & 1 & 1\\
1 & 2 & 1\\
1 & 1 & 3\\
\end{pmatrix}.
\end{align}

\subsection{Mixture of one-component bosons and two-component fermions}\label{b1f2}

\begin{figure}[t]
  \includegraphics[height=1.75in,width=3.35in]{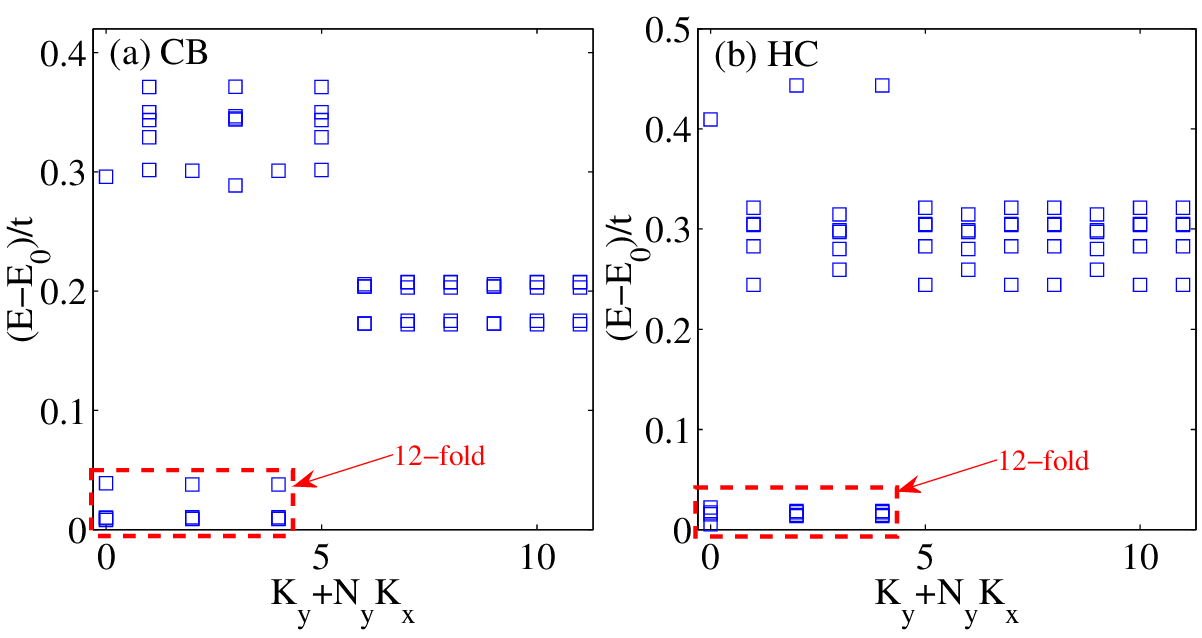}
  \caption{\label{energy2331}(Color online) Numerical ED results for the low energy spectrum of one-component hardcore bosons at $\nu_{+1}=1/3$ and two-component fermions at $\nu_{-1}=\nu_{0}=1/6$ with $U=\infty,V_{0}/t=V_{-1}/t=100,V_{+1}/t=0,$. The system size $N_s=2\times2\times6$ is used for (a) topological checkerboard lattice and (b) topological honeycomb lattice. The lowest five energy levels in each momentum sector are shown. The red dashed box indicates the ground state degeneracy.
  }
\end{figure}

We further examine a mixture of one-component bosons and two-component fermions in the model Hamiltonian, where the particles of spin-$\sigma=+1$ component are taken as spinless hardcore bosons while the particles of spin-$\sigma=0,-1$ component are treated as spinful fermions. We would consider the FQH states of one-component bosons at filling $\nu_{+1}=1/3$ and two-component fermions at fillings $\nu_0=\nu_{-1}=1/6$ with strong onsite repulsions $U/t\gg1$ among different components and only intracomponent nearest-neighboring repulsion between spinful fermions (namely $V_0/t=V_{-1}/t\gg1$ and $V_{+1}=0$). For small system sizes, our ED study of the low energy spectrum confirms a twelve-fold quasi-degenerate ground state manifold, as shown in Figs.~\ref{energy2331}(a) and~\ref{energy2331}(b).

\begin{figure}[t]
  \includegraphics[height=2.5in,width=3.35in]{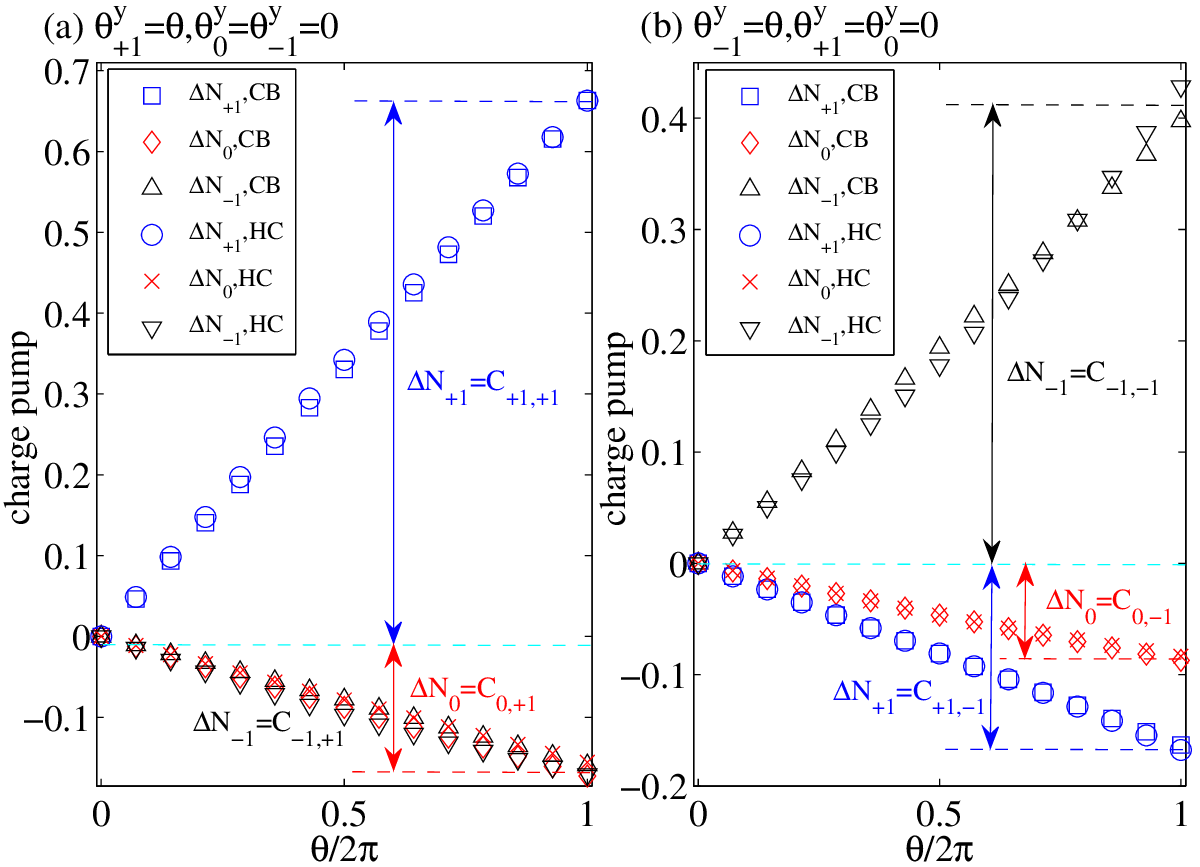}
  \caption{\label{pump233} (Color online) Fractional charge transfers for one-component hardcore bosons at $\nu_{+1}=1/3$ and two-component fermions at $\nu_{-1}=\nu_{0}=1/6$ with $U=\infty,V_{0}/t=V_{-1}/t=100,V_{+1}/t=0,$ on the cylinder lattice for two different topological models under the insertion of two types of flux quantum: (a) $\theta_{+1}^{y}=\theta,\theta_{0}^{y}=\theta_{-1}^{y}=0$ and (b) $\theta_{-1}^{y}=\theta,\theta_{+1}^{y}=\theta_{0}^{y}=0$. Here finite DMRG is used with cylinder width $N_y=4$ and length $N_x=42$.}
\end{figure}

For larger system sizes, as shown in Figs.~\ref{pump233}(a) and~\ref{pump233}(b), we plot the evolution of charge transfers for different types of flux insertion in different topological lattice models, and find that (i) for the flux insertion of spin-$\sigma=+1$ bosons $\theta_{+1}^y=\theta,\theta_0^y=\theta_{-1}^y=0,\theta\subseteq[0,2\pi]$, the charge pumpings
\begin{align}
  &\Delta N_{+1}=N_{+1}^R(2\pi)-N_{+1}^R(0)\simeq C_{+1,+1}=\frac{2}{3}, \nonumber\\
  &\Delta N_{0}=N_{0}^R(2\pi)-N_{0}^R(0)\simeq C_{0,+1}=-\frac{1}{6}, \nonumber\\
  &\Delta N_{-1}=N_{-1}^R(2\pi)-N_{-1}^R(0)\simeq C_{-1,+1}=-\frac{1}{6}. \nonumber
\end{align}
and (ii) for the flux insertion of spin-$\sigma=-1$ fermions, the charge pumpings $\theta_{+1}^y=\theta_0^y=0,\theta_{-1}^y=\theta,\theta\subseteq[0,2\pi]$,
\begin{align}
  &\Delta N_{+1}=N_{+1}^R(2\pi)-N_{+1}^R(0)\simeq C_{+1,-1}=-\frac{1}{6}, \nonumber\\
  &\Delta N_{0}=N_{0}^R(2\pi)-N_{0}^R(0)\simeq C_{0,-1}=-\frac{1}{12}, \nonumber\\
  &\Delta N_{-1}=N_{-1}^R(2\pi)-N_{-1}^R(0)\simeq C_{-1,-1}=\frac{5}{12}. \nonumber
\end{align}
As above we can obtain the $\mathbf{K}$-matrix from the inverse of the Chern number matrix, namely
\begin{align}
  \mathbf{K}=\mathbf{C}^{-1}=\begin{pmatrix}
2 & 1 & 1\\
1 & 3 & 1\\
1 & 1 & 3\\
\end{pmatrix}.
\end{align}
The determinant of this $\mathbf{K}$-matrix also matches with the topological degeneracy of the ground
state manifold in our ED study correspondingly.

\section{Non-Abelian Spin-Singlet Fractional Quantum Hall Effect}\label{nonabelian}

\begin{figure}[t]
  \includegraphics[height=2.2in,width=3.35in]{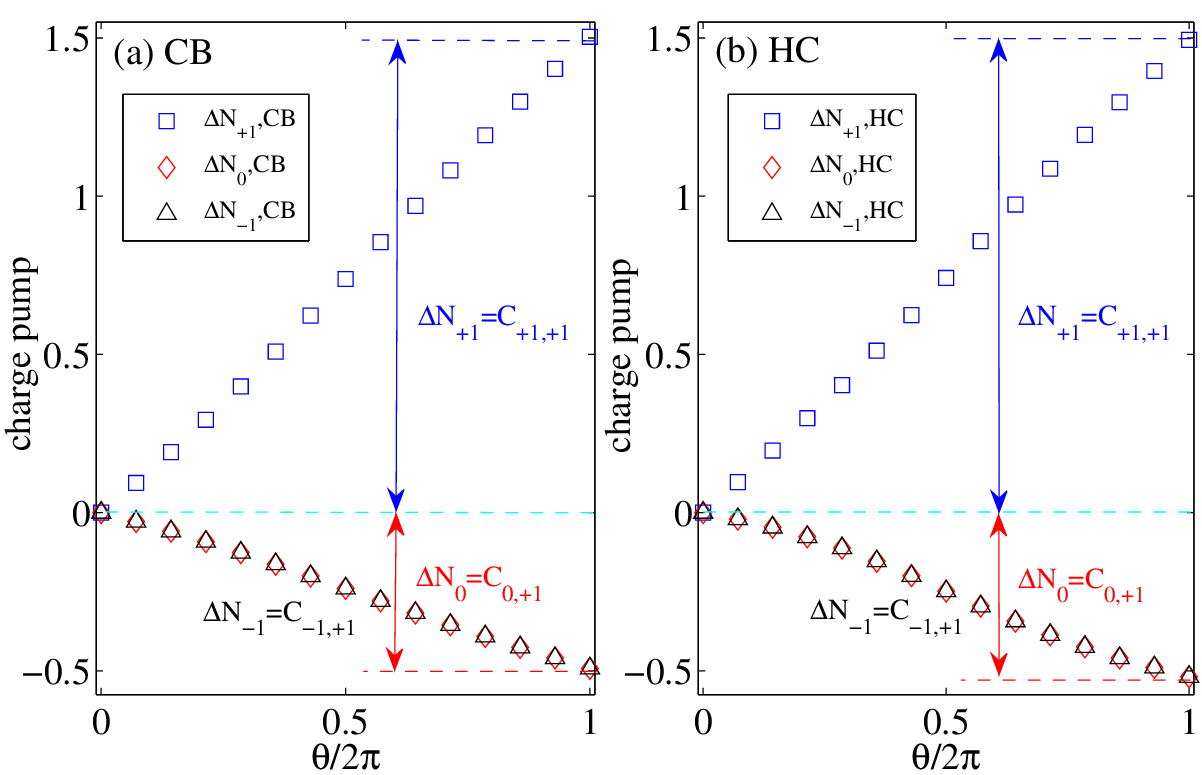}
  \caption{\label{nass} (Color online) Fractional charge transfers for three-component softcore bosons at $\nu=3/2$ with infinite three-body repulsion $U_{3b}=\infty$ on the cylinder lattice under the insertion of flux quantum $\theta_{+1}^{y}=\theta,\theta_{0}^{y}=\theta_{-1}^{y}=0$ for two different topological models: (a) checkerboard model and (b) Haldane-honeycomb model. Here finite DMRG is used with the cylinder width $N_y=3$, keeping the cylinder length up to 36.}
\end{figure}

In this section, we extend our technique to identify non-Abelian SU(3) spin-singlet FQH states for three-component softcore bosons at filling $\nu=3k/4$ in the presence of $(k+1)$-body interactions as a primary example~\cite{Fuji2017}. Following Ref.~\cite{Ardonne1999}, we can construct the trial wavefunction (apart from a constant conformal product factor) from that of Abelian SU(3) FQH state at $\nu=3/4$ as
\begin{align}
 \Psi_{\nu=3k/4} &\propto [\Psi_{\nu=3/4}]^{1/k}  \nonumber\\
 &\propto \prod_{i<j,\sigma}(z_i^{\sigma}-z_j^{\sigma})^{2/k} \prod_{i,j,\sigma\neq\sigma'}(z_i^{\sigma}-z_j^{\sigma'})^{1/k}.\label{waveNASS}
\end{align}
Akin to two-component non-Abelian spin-singlet states as discussed in Ref.~\cite{Zeng2022b}, we claim a peculiar $3\times3$ Chern-number matrix, in connection with the power-law exponents of mutual particle correlations in Eq.~\ref{waveNASS},
\begin{align}
  \mathbf{C}=\frac{k}{4}\begin{pmatrix}
3 & -1 & -1 \\
-1 & 3 & -1 \\
-1 & -1 & 3 \\
\end{pmatrix}=\begin{pmatrix}
2/k & 1/k & 1/k \\
1/k & 2/k & 1/k \\
1/k & 1/k & 2/k \\
\end{pmatrix}^{-1}\label{chernNASS}
\end{align}
Numerically, we here mainly discuss the simplest case of $k=2$ with SU(3) symmetric three-body repulsion in Eqs.~\ref{cbl} and~\ref{hcl},
\begin{align}
  V_{int}=U_{3b}\sum_{\rr}\prod_{i=0}^{k=2}(n_{\rr,+1}+n_{\rr,0}+n_{\rr,-1}-i)
\end{align}
In the strongly interacting regime $U_{3b}=\infty$ (namely no more than $k=2$ particles are allowed per lattice site), our DMRG simulation of the charge pumping on finite cylinder ladders by inserting one flux quantum $\theta_{+1}^y=\theta,\theta_0^y=\theta_{-1}^y=0,\theta\subseteq[0,2\pi]$, as shown in Figs.~\ref{nass}(a) and~\ref{nass}(b), gives
\begin{align}
  &\Delta N_{+1}=N_{+1}^R(2\pi)-N_{+1}^R(0)\simeq C_{+1,+1}=\frac{3}{2}, \nonumber\\
  &\Delta N_{0}=N_{0}^R(2\pi)-N_{0}^R(0)\simeq C_{0,+1}=-\frac{1}{2}, \nonumber\\
  &\Delta N_{-1}=N_{-1}^R(2\pi)-N_{-1}^R(0)\simeq C_{-1,+1}=-\frac{1}{2}, \nonumber
\end{align}
which coincide with the prediction of Eq.~\ref{chernNASS}. Thus, our formulation of the Chern number matrix faithfully manifests the internal structure of non-Abelian multicomponent FQH state here.

As an extension to three-component fermionic non-Abelian spin-singlet states at filling $\nu=3/4$, similar to the single-component Pfaffian state, we expect that there may be two plausible non-Abelian candidates: (a) intercomponent Pfaffian state $\Psi_{\text{inter}}=\prod_{\sigma\neq\sigma'}\text{Pf}(\frac{1}{z_i^{\sigma}-z_j^{\sigma'}})[\Psi_{\nu=3/4}]$ and (b) intracomponent Pfaffian state $\Psi_{\text{intra}}=\prod_{\sigma}\text{Pf}(\frac{1}{z_i^{\sigma}-z_j^{\sigma}})[\Psi_{\nu=3/4}]$, both of which also share the same Chern number matrix as that of bosonic spin-singlet FQH state $\Psi_{\nu=3/4}$ at filling $\nu=3/4$.

\section{Conclusion}\label{summary}

To summarize our findings, we have developed a controllable scheme for studying three-component FQH effects emerging in paradigmatic topological lattice models under the interplay of band topology, intercomponent and intracomponent correlations. We numerically uncover the $\mathbf{K}$-matrix classifications at filling $\nu=3/4$ for three-component hardcore bosons and at various fillings for three-component Bose-Fermi mixtures. We implement this by providing two pieces of major ingredients: (i) $\det|\mathbf{K}|$-fold ground state manifold equivalent to the determinant of the $\mathbf{K}$ matrix from ED calculation, (ii) topological Chern number matrix equivalent to the inverse of the $\mathbf{K}$ matrix from DMRG calculation. We also propose a Chern number matrix to characterize non-Abelian three-component spin-singlet FQH effect. According to the $\mathbf{K}$-matrix theory~\cite{Wen1995}, the quasiparticle charge is given by $Q_q=\mathbf{t}^{T}\cdot\mathbf{K}^{-1}\cdot\mathbf{l}$ where $\mathbf{t}=(t_{+1},t_0,t_{-1})^T$ is the charge vector and $\mathbf{l}=(l_{+1},l_0,l_{-1})^T$ is the quasiparticle charge vector. For a generic quasiparticle excitation labeled by the $\mathbf{l}$ vector, consists of $l_{\sigma}$ number of quasiparticles of $\sigma$-th component. In connection to our model, each component (spin $\sigma=0,\pm1$) contributes the Hall transport, and the charge vector $\mathbf{t}=(1,1,1)^T$. Also the quasiparticle charge vector can be $\mathbf{l}=(1,1,1)^T$, and finally, the quasiparticle charge $Q_q=\sum_{\sigma,\sigma'}C_{\sigma,\sigma'}$, consists of charge $Q_{\sigma}=\sum_{\sigma'}C_{\sigma,\sigma'}$ of $\sigma$-th component, which just equal to the charge transfer in our DMRG simulation. It is noteworthy that our comprehensive DMRG simulation of fractional charge pumping of multicomponent systems reveals a wealth of crucial information about the topological order, ultimately serving as a practical tool for probing exotic quantum Hall physics.

\begin{acknowledgements}
T.S.Z thanks D. N. Sheng and W. Zhu for inspiring discussions and prior collaborations on multicomponent fractional quantum Hall physics in topological flat band models.
This work is supported by the National Natural Science Foundation of China (NSFC) under Grant No. 12074320. We also acknowledge the ALPS open source package for implementing the DMRG algorithm~\cite{Dolfi2014}.
\end{acknowledgements}

\appendix

\section{Topological lattice geometry}\label{band}

\begin{figure}[b]
  \centering
  \includegraphics[height=1.5in,width=2.9in]{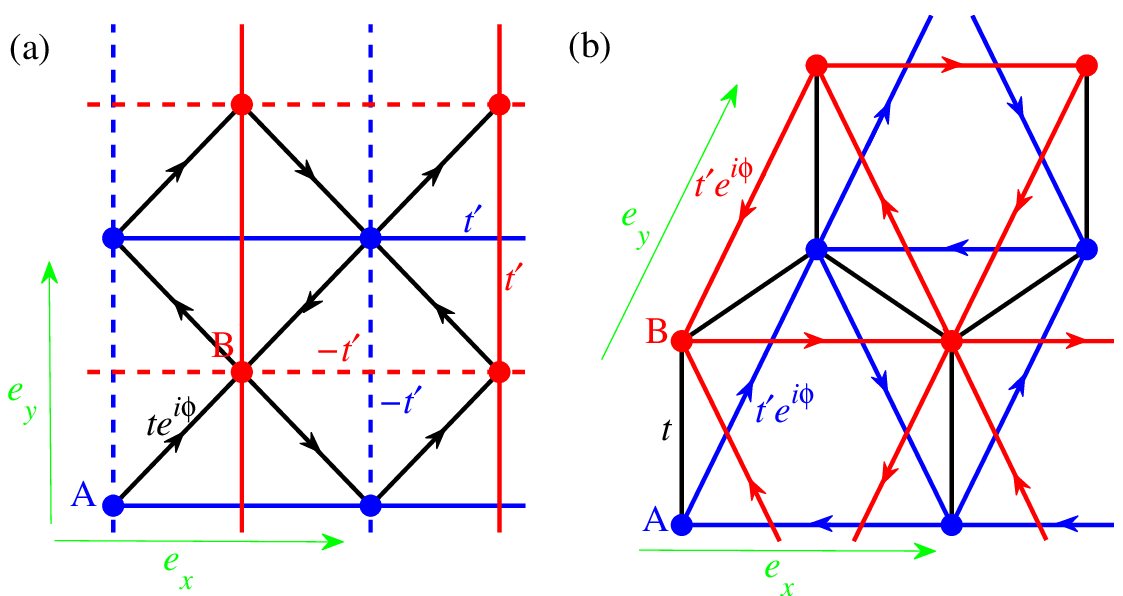}
  \includegraphics[height=1.5in,width=2.9in]{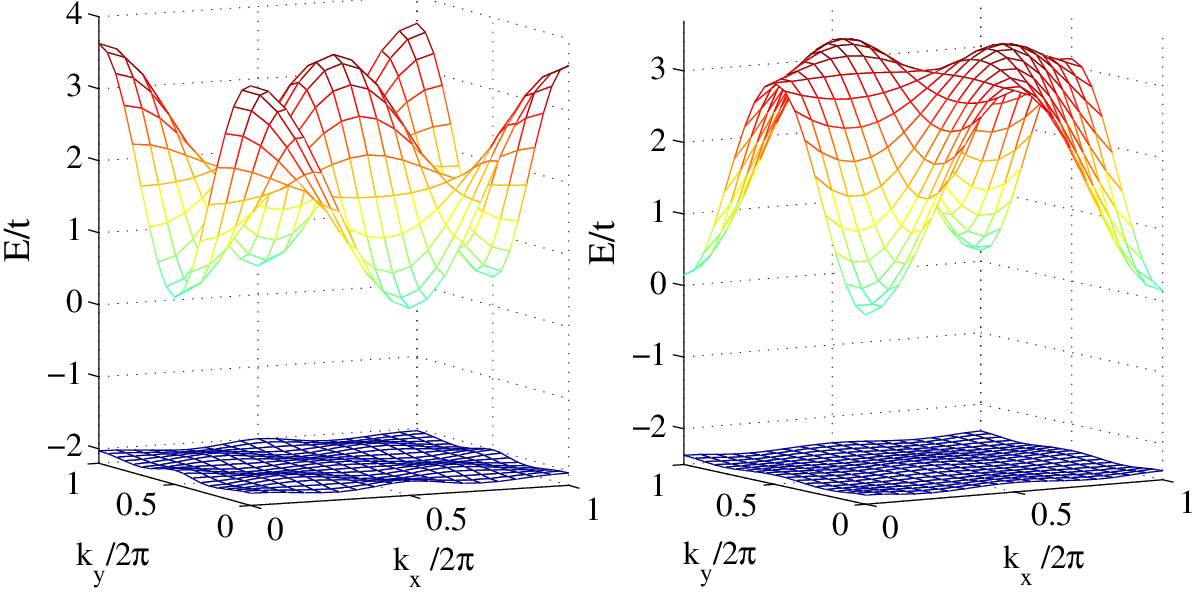}
  \caption{\label{lattice}(Color online) The schematic plot of (a) $\pi$-flux checkerboard lattice model and (b) Haldane-honeycomb lattice model and the corresponding band structure.}
\end{figure}

Here we append the description of the topological lattice geometry with the hopping parameters in the main text. As indicated in Figs.~\ref{lattice}(a) and~\ref{lattice}(b), two inequivalent sublattice sites $A,B$ are labeled by blue (red) solid circles. The chiral flux phase $\phi_{\rr'\rr}=\phi$ with a positive sign along the hopping direction is marked by the arrow link. For the checkerboard lattice, the next-nearest-neighbor hopping amplitudes are $t_{\rr,\rr'}'=\pm t'$ along the solid (dashed) lines. The real-space lattice translational vectors are indicated by $e_{x,y}$. The bottom panels show the corresponding band dispersions. For tunnel couplings $t'=0.3t,t''=-0.2t,\phi=\pi/4$, the band width $W$ of the lowest Chern band is about $W\simeq0.1t$ for checkerboard lattice, while for $t'=0.6t,t''=-0.58t,\phi=2\pi/5$, the band width $W$ of the lowest Chern band is about $W\simeq0.05t$ for honeycomb lattice, such that the band width is much smaller than the band gap and interaction energy scales, and the stability of FQH states is enhanced.

\section{Chern number matrix}\label{chern}

In interacting systems, topological Chern number of the many-body ground state can be obtained using twisted boundary conditions $c_{\rr+N_{\alpha},\sigma}=c_{\rr,\sigma}\exp(i\theta_{\sigma}^{\alpha})$ in the $\alpha=x,y$ direction where $\theta_{\sigma}^{\alpha}$ is the twisted angle which leads to a shift of lattice momentum $k_{\alpha}\rightarrow k_{\alpha}+\theta_{\sigma}^{\alpha}/N_{\alpha}$ in the spin-$\sigma$ component~\cite{Niu1985}. In the parameter plane of two independent twisted angles $\theta_{\sigma}^{x}\subseteq[0,2\pi],\theta_{\sigma'}^{y}\subseteq[0,2\pi]$, the Chern number of the many-body ground state wavefunction $\psi(\theta_{\sigma}^{x},\theta_{\sigma'}^{y})$ is defined by the integral formula $C_{\sigma\sigma'}=\int\int d\theta_{\sigma}^{x}d\theta_{\sigma'}^{y}F_{\sigma\sigma'}(\theta_{\sigma}^{x},\theta_{\sigma'}^{y})/2\pi$, where the Berry curvature $F_{\sigma\sigma'}(\theta_{\sigma}^{x},\theta_{\sigma'}^{y})=\mathbf{Im}\left(\langle{\frac{\partial\psi}{\partial\theta_{\sigma}^x}}|{\frac{\partial\psi}{\partial\theta_{\sigma'}^y}}\rangle
-\langle{\frac{\partial\psi}{\partial\theta_{\sigma'}^y}}|{\frac{\partial\psi}{\partial\theta_{\sigma}^x}}\rangle\right)$.
For three-component systems, we can write nine elementary Chern numbers $C_{\sigma\sigma'}$ for $\sigma,\sigma'=0,\pm1$, which constitute a Chern number matrix spanned in spin space~\cite{Sheng2003,Sheng2006}
\begin{align}
  \mathbf{C}=\begin{pmatrix}
C_{+1,+1} & C_{+1,0} & C_{+1,-1}\\
C_{0,+1} & C_{0,0} & C_{0,-1}\\
C_{-1,+1} & C_{-1,0} & C_{-1,-1}\\
\end{pmatrix}.
\end{align}

\section{Chiral edge modes}

\begin{figure}[t]
  \includegraphics[height=1.8in,width=2.8in]{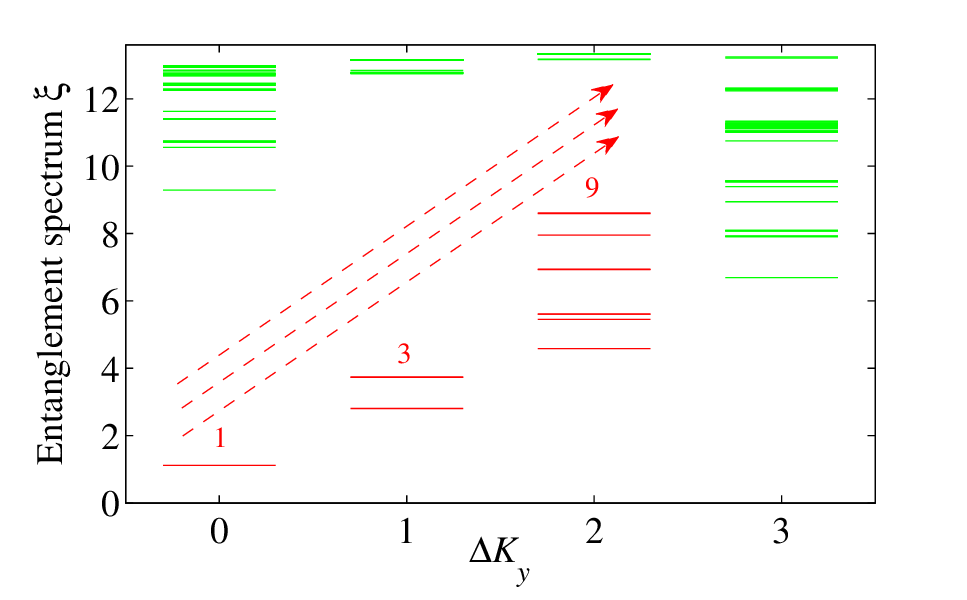}
  \caption{\label{es} (Color online) Chiral edge mode identified from the momentum-resolved entanglement spectrum for three-component hardcore bosons at $\nu=3/4$ with $U=\infty,V_{\sigma}/t=0,$ on the cylinder of topological honeycomb lattice. The horizontal axis shows the relative momentum $\Delta K_y=K_y-K_{y}^{0}$ (in units of $2\pi/N_y$). Numbers below the dashed red line label the excitation level counting $1,3,9$ at different momenta for a typical charge sector $\Delta Q_{+1}=\Delta Q_{0}=\Delta Q_{-1}=0$. Here infinite DMRG is used with the maximal bond dimension kept up to $2000$. }
\end{figure}

Indeed, according to the Chern-Simons theory, the chirality of edge modes is determined by the signs of the eigenvalues of the $\mathbf{K}$ matrix~\cite{Wen1995}. For the present Abelian FQH states in the main text, we can determine that all of the eigenvalues of the $\mathbf{K}$ matrix host positive signs, implying three propagating chiral modes in the same direction. By diagonalizing the three-component Lagrangian in the effective field theory similar to the procedure in Ref.~\cite{Furukawa2013}, we expect that given the charge sectors, the edge Hamiltonian and the corresponding momentum operator are (apart from certain constant shift)
\begin{align}
  H_{\text{edge}}&=\frac{2\pi}{N_y}(v_1L_1+v_2L_2+v_3L_3) \\
  \Delta K_y&=\frac{2\pi}{N_y}(L_1+L_2+L_3) \\
  L_{i}&=\sum_{k=1}^{\infty}kn_{k}^i, \quad i=1,2,3 \nonumber
\end{align}
where $v_i>0$ is the chiral propagating velocity (Approximately, $v_1\simeq v_2\simeq v_3$), and $\{n_k^i\}$ denotes the set of non-negative integers of $i$-th edge branch excitation mode. We can simulate the excitation level in momentum space, and derive the degenerate level patterns of $\{n_k^i\}$ for any momentum. For example, the zeroth excited level corresponds to $\Delta K_y=0$ with only one pattern $n_k^i=0$; the first excited level corresponds to $\Delta K_y=1$ with three degenerate patterns: (1) $n_{k=1}^1=1,n_{k\neq1}^1=0,n_k^{i\neq1}=0$, (2) $n_{k=1}^2=1,n_{k\neq1}^2=0,n_k^{i\neq2}=0$, and (3) $n_{k=1}^3=1,n_{k\neq1}^3=0,n_k^{i\neq3}=0$. And the corresponding excitation level counting for three free bosonic edge modes is given by
\begin{align}
  \Delta K_y&: \quad 0 \quad\quad 1 \quad\quad 2 \quad\quad 3 \nonumber\\
  \text{Degeneracy}&: \quad 1 \quad\quad 3 \quad\quad 9 \quad\quad 22 \nonumber
\end{align}

Numerically, we calculate the low-lying momentum-resolved entanglement spectrum on the cylinder using infinite DMRG. The infinite cylinder is divided into left-half and right-half parts along the $x$ direction, and the entanglement spectrum is defined as $\xi_i=-\ln\rho_i$ where $\rho_i$ is the eigenvalue of the density matrix $\widehat{\rho}_L$ of the left-half cylinder, which would reveal the characteristic chirality and excited level counting of edge modes~\cite{Li2008}. As an example, we consider the FQH state of three-component hardcore bosons at filling $\nu=3/4$, and plot the low-lying momentum-resolved entanglement spectrum on the infinite cylinder with width $N_y=4$. As indicated in Fig.~\ref{es}, we observe the low-lying level counting $1,3,9,\cdots$ separated by a large entanglement gap from the higher levels, consistent with theoretical analysis of three forward-moving edge branches in the same direction, as discussed above.

%


\end{document}